\documentclass[aps,prl,showpacs,twocolumn,superscriptaddress]{revtex4}%
\usepackage{epsfig,dsfont,amssymb,amsmath,amsthm,amsfonts,amsbsy,mathrsfs}
\usepackage{graphicx}
\usepackage{graphicx}
\usepackage{color}
\usepackage{ulem}
\usepackage{lipsum}

\begin{document}

\title{Inhomogeneous screening near a dielectric interface}

\author{Rui Wang}

\author{Zhen-Gang Wang}
\email {zgw@caltech.edu}
\affiliation{Division of Chemistry and Chemical Engineering,California Institute of Technology, Pasadena, CA 91125, USA}


\begin{abstract}

Screening is one of the most important concepts in the study of charged systems.  Near a dielectric interface, the ion distribution in a salt solution
can be highly nonuniform.  Here, we develop a theory that
self-consistently treats the inhomogeneous screening effects.  At
higher concentrations when the bulk Debye screening length is
comparable to the Bjerrum length, the double layer structure and
interfacial properties are significantly affected by the
inhomogeneous screening. In particular, the depletion zone is
considerably wider than that predicted by the bulk screening
approximation or the WKB approximation.  For asymmetric salts, the
inhomogeneous screening leads to enhanced charge separation and
surface potential.

\end{abstract}


\maketitle

Screening due to the ionic atmosphere, introduced by Debye and
H\"uckel more than 90 years ago\cite{Debye}, is one of the most
important concepts in the study of charged systems.  Screening has
profound effects on essentially all properties of biophysical and
salt-containing soft matter
systems\cite{Israelachvili,McQuarrie,Andelmanreview,Levinreview,Hansen,Honig,Gelbart}.
In a homogeneous bulk solution, screening is most commonly
manifested as an exponential damping of the long-range Coulomb
interactions between two test charges.  When the ion distribution is
nonuniform, as in the vicinity of a charged surface or an interface
with dielectric discontinuity, screening also becomes inhomogeneous.
However, in spite of the ubiquity of systems with nonuniform ion distributions, a rigorous treatment of screening in such systems is still lacking.

For a salt solution near a dielectric interface, e.g., the water/air
interface, the repulsive image force creates a depletion layer,
whose theoretical treatment was pioneered by Onsager and Samaras
(OS) \cite{onsager}. This problem is related to a number of phenomena, such as
conductivity in artificial and biological
ion-channels\cite{Channel1,Channel2,Channel3,Channel4},
stability of colloidal, bubble and protein
suspensions\cite{Stability1,Stability2,Stability3,Stability4}, and
the rate of ozone consumption\cite{Ozone1,Ozone2}.  Assuming that the
image force is screened by the bulk screening length, the OS theory
qualitatively explains the excess surface tension and yields
agreement with experimental data at low salt concentrations ($c_b<0.01M$).  However, there is large discrepancy between
the OS theory and experiment data at high salt concentrations
($c_b>0.1M$)\cite{surfacetension1,surfacetension2}.  The OS theory
predicts an ever decreasing width of the depletion layer with salt
concentration, which results in a concave downwards curve
for the surface tension vs. $c_b$.  In contrast, experimental data show
essentially a linear increase of the surface tension.  To reconcile
this discrepancy, an exclusion zone of constant width or large
hydration radius of the ion is usually invoked\cite{Levin1,Levin2}.

\begin{figure}[bht]
\centering
\includegraphics[width=0.35\textwidth]{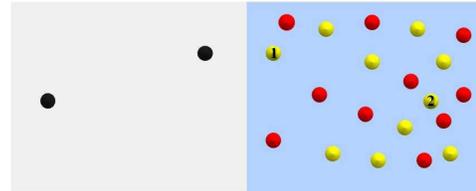}
\caption{Schematic of the inhomogeneous screening near the
dielectric interface. The red and yellow spheres represent the
cations and anions, respectively. The two test ions are labeled,
with Ion 1 located very close to the interface and Ion 2 approaching
the bulk solution. The two black spheres are the image charges
corresponding to the two test ions.}
\end{figure}

An obvious effect missing in the OS theory\cite{onsager} and in subsequent
modifications\cite{Levin1,Levin2,Levin3,Monica,Bell2,Ninham,Parsons,Onuki1,Onuki2,Buyukdagli,Dean,Markovich}
is the spatially varying screening of the image force near a dielectric interface (see Fig. 1): the ion concentration changes
gradually from zero at the interface to the bulk value. In this
depletion layer, the ionic cloud is highly anisotropic, giving rise to different features of the screening near
the interface from the homogeneous and isotropic bulk. Close to the
interface, ions are strongly depleted; the local ionic strength
around the test ion (Ion 1 in Fig. 1) is much lower than the bulk.
The bulk screening approximation clearly overestimates the screening
effect and hence underestimates the image force.  Even for an ion
approaching the bulk (Ion 2 in Fig. 1), the screening is still
weaker than in the bulk due to the long-range, accumulative
effects from the depletion zone.  This
feature extends the effective range of the image force beyond the
Debye screening length.  The WKB
approximation\cite{Buff} provides an approximate treatment of the inhomogeneous nature of screening by using the local ionic strength; however, it does not capture the long-range,  accumulative effects.

Since approximate treatments cannot fully account for all
the features of inhomogeneous screening,  previous calculations using these approximations for the double layer structure and interfacial properties are likely to be inaccurate.  Such inaccuracy in treating the essential electrostatic contributions makes it impossible to evaluate the relative importance of the various non-electrostatic effects invoked, for example, to
explain the surface tension behavior, such as the cavity
energy\cite{Levin1,Levin2},  hydration\cite{Levin2}, and
dispersion forces\cite{Ninham,Parsons}.  In this Letter, we examine
the issue of inhomogeneous screening in salt solutions near a dielectric
interface, by comparing the result from numerical solution of the full
Green function with results obtained using approximate methods. We
find that the effects of inhomogeneous screening on the double
layer structure and interfacial properties are quite pronounced as
the Debye screening length becomes comparable to the Bjerrum length.

We have recently developed a general theory for weak-coupling
systems with a fixed charge distribution $\rho_{ex}$ in the presence
of mobile cations with charge $q_+ e$ and anions with charge $q_-
e$, in a dielectric medium of a spatially varying dielectric
function $\varepsilon $\cite{Wangzg,wang2}.  The key result is the
set of self-consistent equations for the mean electrostatic
potential $\psi({\bf r})$ (nondimensionized by $kT/e$), the Green
function $G({\bf r}, {\bf r}')$, and the self-energy $u_{\pm} ({\bf
r})$ of the ions:
\begin{equation}
- \nabla \cdot \left( \epsilon \nabla \psi \right) = \rho_{ex} +
\Gamma  \lambda_+  q_+ {\rm e}^{ - q_+  \psi -  u_+ } - \Gamma
\lambda_- q_- {\rm e}^{q_-  \psi - u_- } \label{eq1}
\end{equation}
\begin{equation}
- \nabla \cdot \left[ \epsilon \nabla G({\bf r},{\bf r}') \right] +
2  I({\bf r}) G({\bf r},{\bf r}')=  \delta ({\bf r}-{\bf r}')
\label{eq2}
\end{equation}
\begin{equation}
u_{\pm} ({\bf r})= \frac{1}{2}  \int d {\bf r}' d {\bf r}'' h_{\pm}
({\bf r}-{\bf r}') G({\bf r}',{\bf r}'') h_{\pm} ({\bf r}''-{\bf r})
\label{eq3}
\end{equation}
where $\epsilon=kT \varepsilon_0 \varepsilon /e^2$ is the scaled
permittivity and $\lambda_{\pm}$ is the fugacity of ions. $\Gamma$
is introduced to constrain the mobile ions to the solvent region.
$I({\bf r})=\left[q_+^2 c_+({\bf r})+q_-^2 c_-({\bf r})\right]/2$ is
the local ionic strength, with the ion concentration given by
$c_{\pm} ({\bf r}) =\lambda_{\pm} \Gamma \exp \left[ \mp q_{\pm}
\psi ({\bf r}) -u_{\pm} ({\bf r}) \right]$. The short-range charge
distribution $h_{\pm} ({\bf r}-{\bf r}')$ on the ion is introduced
to yield a finite Born solvation energy. Eq. \ref{eq1} is the
self-energy modified Poisson-Boltzmann equation, reflecting that the
ion distribution is determined by both the mean electrostatic
potential and the self energy. From Eqs. \ref{eq2} and \ref{eq3},
the inhomogeneity in the ionic strength affects the solution of the
Green function and the self energy, which consequently affect the
double layer structure through Eq. \ref{eq1}.

We now specify to a salt solution in contact with a low dielectric
medium through a sharp interface (at $z=0$) with fixed surface
charge density $\rho_{ex}({\bf r})=\sigma(z)$. Mobile ions are
excluded from the low dielectric side. Both $\Gamma$ and
$\varepsilon$ are then step functions: $\Gamma=0$ and
$\varepsilon=\varepsilon_{P}$ for $z<0$; $\Gamma=1$ and
$\varepsilon=\varepsilon_{S}$ for $z>0$. In the solvent region
($z>0$), Eq. \ref{eq1} becomes
\begin{equation}
-\epsilon_{S} \frac{\partial^2 \psi (z)}{\partial z^2}= \lambda_+
q_+ {\rm e}^{ - q_+  \psi -  u_+ } -  \lambda_-  q_- {\rm e}^{q_-
\psi - u_- } \label{eq5}
\end{equation}
with boundary condition $(\partial \psi / \partial
z)_{z=0}=-\sigma/\epsilon_{S}$. Assuming the solvent has a uniform
dielectric constant in the entire $z>0$ region, the Born energy is
constant and can be absorbed into the reference chemical potential.
The remaining contribution is finite in the point-charge
limit [$h_{\pm}({\bf r}-{\bf r}')=q_{\pm}\delta({\bf r}-{\bf r}')$],
leading to the nontrivial and nonsingular part of the self energy as
$u_{\pm}^{*}=(q_{\pm}^2 /2)  \lim_{{\bf r}' \to {\bf r}} \left[
G({\bf r}, {\bf r}')- 1/\left( 4 \pi \epsilon_{S} \vert {\bf r} -
{\bf r}' \vert \right) \right] $.

To solve the Green function in the planar geometry, it is convenient
to work in a cylindrical coordinate $(r,z)$ and use the Fourier representation
in the transverse directions:
\begin{equation}
G(r,z,z')=\frac{1}{2\pi} \int _0 ^\infty kdk J_0(k r) {\hat
G}(k,z,z') \label{eq7}
\end{equation}
where $J_0$ is the 0th-order Bessel function.  It is easy to show that ${\hat G}(k,z,z')$
satisfies:
\begin{equation}
 -\frac{\partial^2 {\hat
G}(k,z,z')}{\partial z^2}+\left[ \kappa^2 (z)+k^2 \right] {\hat
G}(k,z,z')=\frac{1}{\epsilon_S} \delta(z,z') \label{eq8}
\end{equation}
for $z>0$, with the boundary condition $\epsilon_S \partial {\hat G}
/
\partial z- k \epsilon_P {\hat
G} =0$ at $z=0$\cite{boundary}.  $\kappa(z)=\left[2 I
(z)/\epsilon_{S}\right]^{1/2}$ can be considered the inverse of the
local Debye screening length.

The bulk screening approximation widely used in the literature \cite{Monica,onsager,Levin1,Levin3,
Levin2,Bell2,Ninham,Parsons,Onuki1,Onuki2,Buyukdagli,Dean,Markovich},
replaces the spatially varying screening length $\kappa(z)$ in
Eq. \ref{eq8} by the constant bulk screening length $\kappa_b$, which enables an analytical solution for the Green function as
${\hat G}(k,z,z')=\left[{\rm e}^{- \omega \vert z-z' \vert} + \Delta
{\rm e}^{- \omega (z+z')} \right]/(2 \epsilon_S \omega)$, where
$\omega=\sqrt{\kappa_b^2+k^2}$ and $\Delta=(\epsilon_S \omega-
\epsilon_P k )/(\epsilon_S \omega+ \epsilon_P k)$. Substituting
${\hat G}(k,z,z')$ into Eq. \ref{eq7} leads to the following
intuitive form for the self energy when $\epsilon_S \gg \epsilon_P$:

\begin{equation}
u_{\pm}^{*}=\frac{q_\pm ^2}{8 \pi \epsilon_{S} }
\left(-\kappa_b+\frac{f {\rm{e}}^{-2 \kappa_b z}}{2z} \right)
\label{eq10}
\end{equation}
with $f=(\epsilon_S - \epsilon_P )/(\epsilon_S + \epsilon_P )$ the
dielectric contrast. The
WKB approximation\cite{Buff,wang1,Levine} is simply to replace the bulk $\kappa_b$ in
Eq. \ref{eq10} by the local $\kappa (z) =\left[2 I
(z)/\epsilon_{S}\right]^{1/2}$.

In this work, we perform the full numerical calculation of the Green
function using the finite difference
method\cite{numerical,Jiaotong}. In order to ensure consistent
numerical accuracy in removing the singularity of the same-point
Green function, the free-space Green function satisfying
$-\partial^2{\hat G}_0/\partial z^2 + k^2 {\hat G}_0 =
\delta(z,z')/\epsilon_S$, is also solved numerically along with Eq.
\ref{eq8}.  $u_{\pm}^{*}$ is then
\begin{equation}
u_{\pm}^{*} (z)=\frac{q_{\pm}^2}{4\pi} \int _0 ^\infty  \left[ {\hat
G}(k,z,z) -  {\hat G}_0(k,z,z)\right] kdk \label{eq11}
\end{equation}
Far away from the interface ($z \to \infty$), the ion concentration
approaches the bulk value $c^b_{\pm}$. It is straightforward to show
$\lambda_\pm=c^b_{\pm} \exp\left[ -q^2_\pm \kappa_b/(8\pi
\epsilon_{S}) \right]$\cite{Wangzg}.

\begin{figure*}[htb]
\centering
\includegraphics[width=0.32\textwidth]{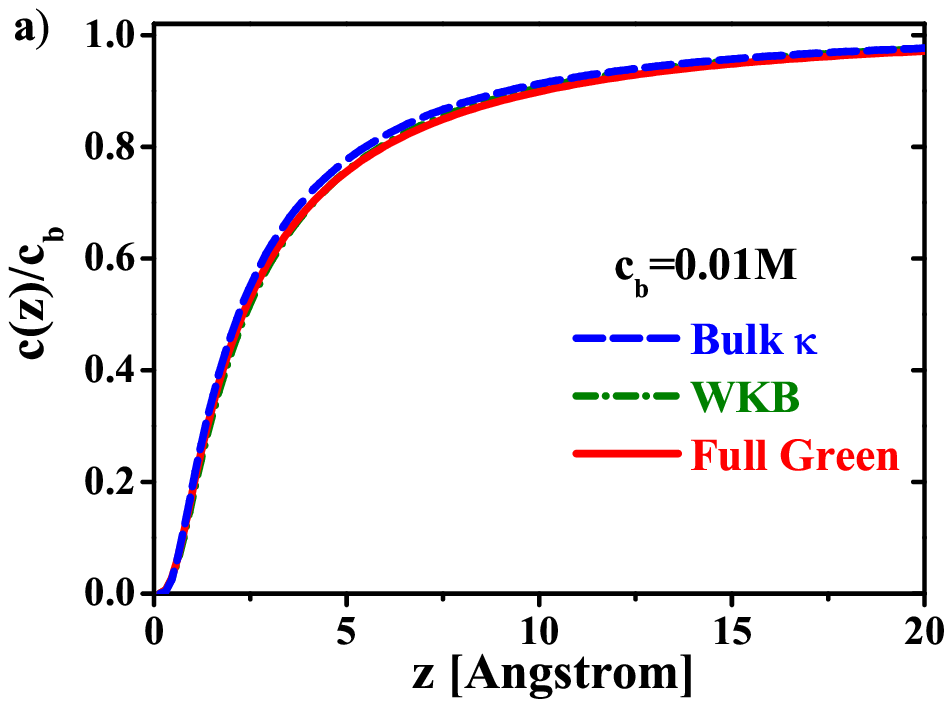}
\includegraphics[width=0.32\textwidth]{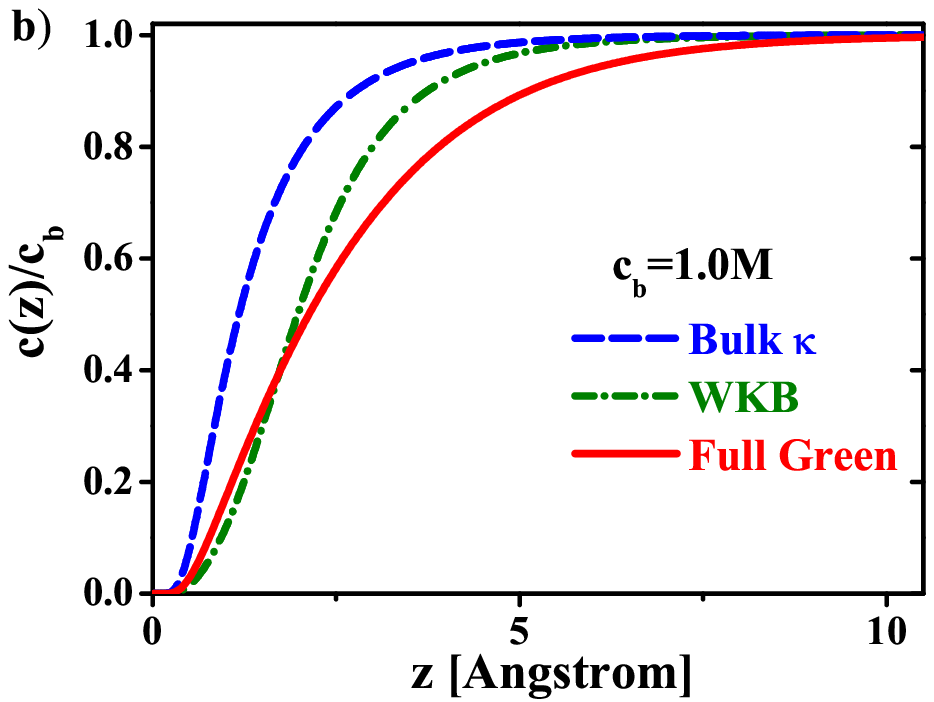}
\includegraphics[width=0.305\textwidth]{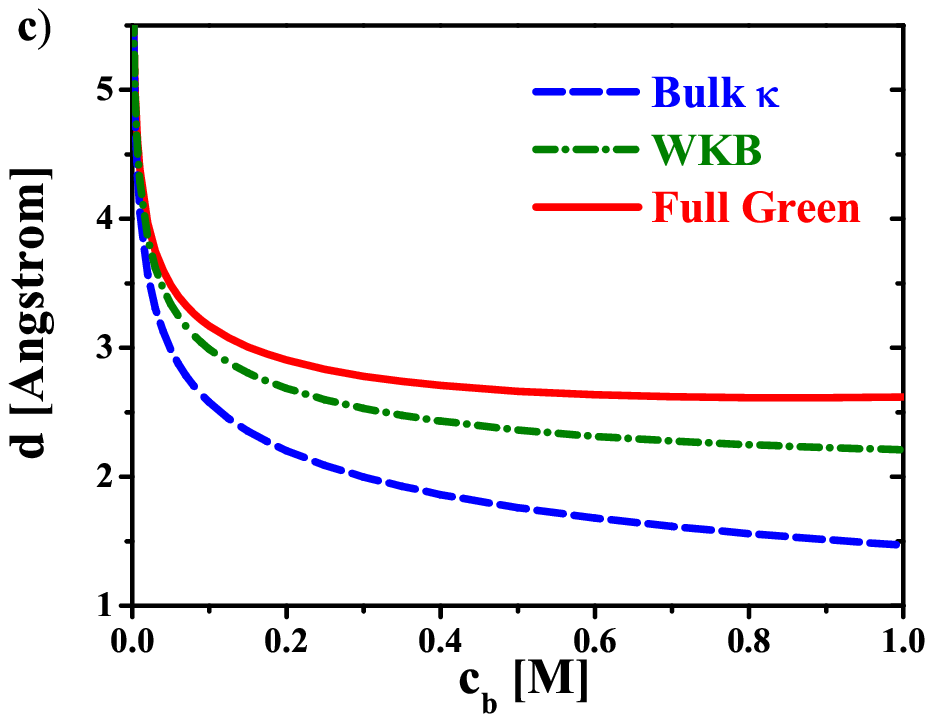}
\caption{Effect of inhomogeneous screening on the ion distribution
of (a)$0.01M$ and (b)$1.0M$ 1:1 salt solution near the water/air
interface. (c) the characteristic length of ion depletion vs. salt
concentration. ``Bulk $\kappa$", ``WKB" and ``Full Green" refer to
the bulk screening approximation, the WKB approximation and
numerically solving the full Green function, respectively.
 \label {2}}
\end{figure*}

\begin{figure}[b]
\includegraphics[width=0.45\textwidth]{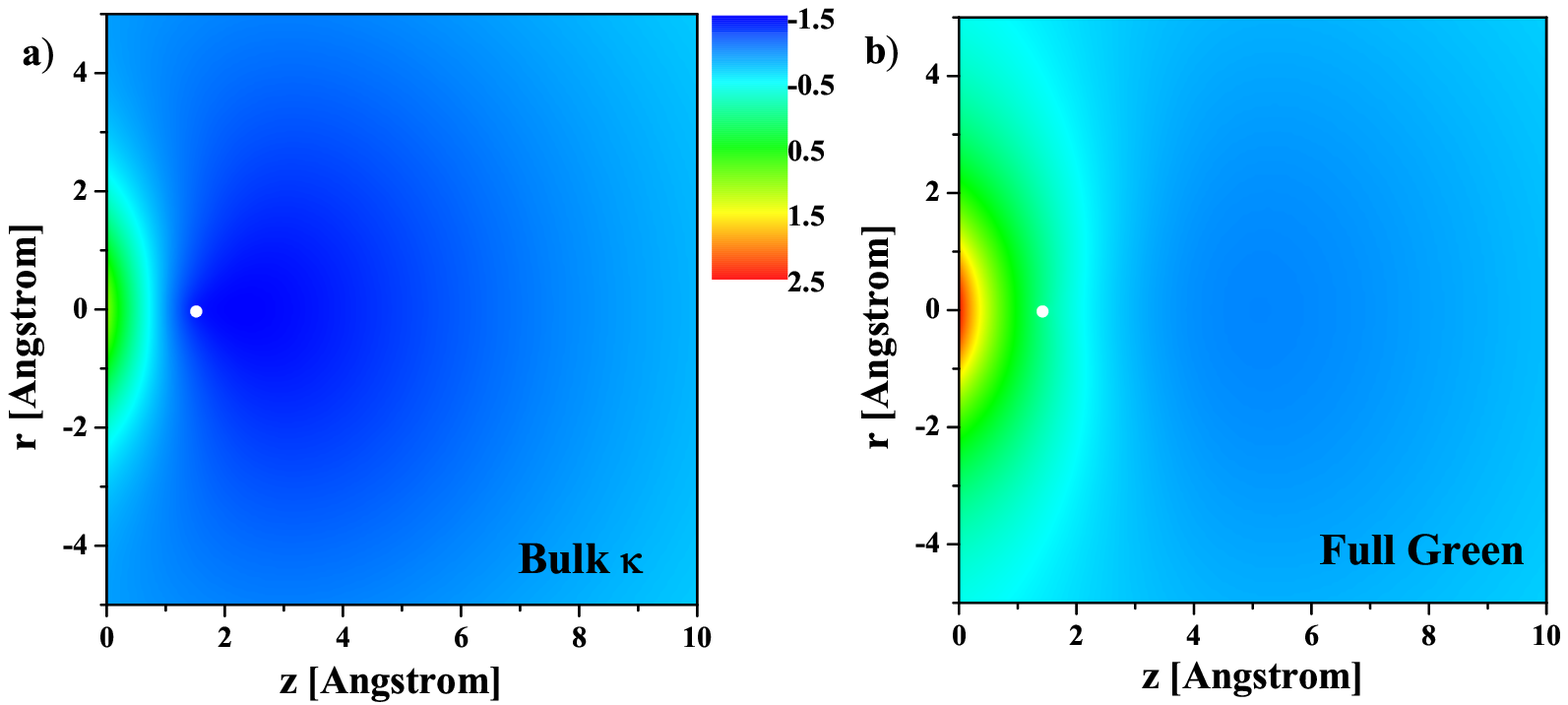}
\includegraphics[width=0.45\textwidth]{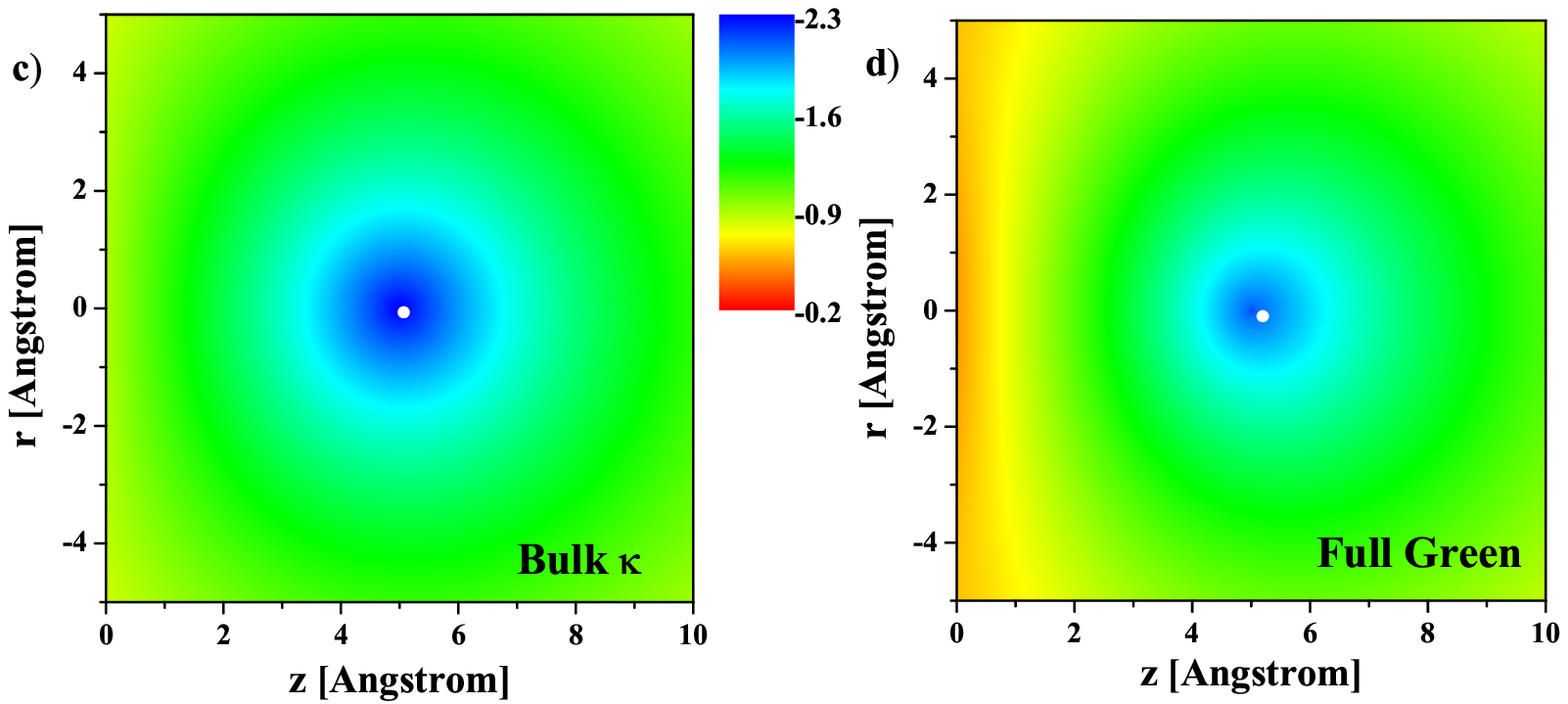}
\caption{2D visualization of the nondivergent Green function,
$G({\bf r}, {\bf r}')- 1/\left( 4 \pi \epsilon_{S} \vert {\bf r} -
{\bf r}' \vert\right)$, for $1.0M$ 1:1 salt solution near the
water/air interface. The test ion (white dot) is at
$z'=1.5\mathring{A}$ for (a) and (b), and $z'=5\mathring{A}$ for (c)
and (d).  \label {2}}
\end{figure}

We now apply the theory to salt solutions near the water/air
interface ($\varepsilon_S=80$ and $\varepsilon_P=1$) with zero fixed
surface charge ($\sigma=0$). This is the same system studied by
Onsager and Samaras.  Figure 2(a) and 2(b) show the concentration profile of the ions in a 1:1 salt solution for two bulk concentrations.
At low salt concentrations when
$\kappa_b^{-1}\gg q^2 l_B$ ($l_B=1/4\pi \epsilon_S$ is the Bjerrum length), the effect of inhomogeneous screening is insignificant, because screening is weak even in the
bulk.  In this regime, both the bulk screening approximation and the WKB approximation are valid.

In contrast, at higher salt concentrations when $\kappa_b^{-1}$
becomes comparable to or even smaller than $q^2 l_B$,  inhomogeneous
screening affects the entire range of the depletion layer as shown
in Fig. 2(b).  Close to the interface ($z<l_B$), the ion
concentration calculated by fully solving the Green function is
significantly lower than that predicted by the bulk screening
approximation, because the local ionic strength that screens the
image force is obviously smaller than the bulk.  In Figs. 3(a) and
3(b) we provide a more visual representation of inhomogeneous
screening by plotting the nondivergent part of the Green function,
$G({\bf r}, {\bf r}')- 1/\left( 4 \pi \epsilon_{S} \vert {\bf r} -
{\bf r}' \vert\right)$, i.e., the nondivergent part of the linear
response electrostatic potential generated by point charge at a
given distance from the interface. The potential generated by the
ion close to the interface ($z=1.5\mathring{A}$) is much stronger
than that predicted by the bulk screening approximation, the latter
severely overestimating the local screening effect on the image
charge interaction. Although this local effect is captured by the
WKB approximation, neither of these two approximations capture the
long-range and accumulative nature of the screening. The depletion
layer calculated by fully solving the Green function extends to a
range significantly longer than the bulk Debye screening length.  As
shown in Fig 3(d), even for an ion approaching the bulk solution
($z=5\mathring{A}$, which is larger than the bulk screening length
of $3.3\mathring{A}$), the electric field from its image charge is
not screened out. This remaining image charge interaction in turn
has a long-range and accumulative effect that reinforces the field
at the position of the point charge.  The two approximate methods
become progressively poorer as the salt concentration increases.

We define $d=\int_0^{\infty} \left[c_b-c(z)\right]dz/c_b$ to
characterize the width of the ion depletion layer, which is shown in
Fig 2(c) as a function of the salt concentration. The two
approximate methods predict $d$ to be an ever decreasing function of
$c^b$, determined by the bulk Debye screening length ($d \sim
\kappa_b^{-1}$). In contrast, $d$ calculated by fully solving the
Green function deviates significantly from the results of the
approximate methods as $\kappa_b^{-1}$ becomes comparable to
$q^{2}l_B$, and reaches a constant value as $c_b$ further increases
up to $1M$. Thus, at high salt concentrations the image-charge
repulsion renormalized by the inhomogeneous screening creates a
depletion layer of nearly constant width scaled by the
Bjerrum length ($d \sim q^2 \l_B$) in stead of $\kappa_b^{-1}$ and becomes nearly independent of the salt concentration.

As a consequence of the different behavior in the width of the
depletion layer due to inhomogeneous screening, there is pronounced
difference in the negative adsorption of ions
($-\Gamma=\int_0^{\infty} \left[c(z)-c_b\right]dz$) at the interface
between results obtained by fully solving the Green function and
those using the approximate methods as shown in Fig 4.  Because the
approximate methods predict an ever decreasing $d$ as $c_b$
increases , $-\Gamma$ is a concave downwards function of $c_b$, as
first shown by Onsager and Samaras\cite{onsager}.  However,
experimentally both $-\Gamma$ and the surface tension of the 1:1
salt solution increases essentially linearly with $c_b$ in the range
of $0.1M<c_b<1M$\cite{surfacetension1,surfacetension2}. To fit the
experimental data, an ion-exclusion zone with constant width has
been invoked in previous theoretical treatments\cite{Levin1,Levin2}.
By fully accounting for the inhomogeneous screening, our theory
naturally predicts that $-\Gamma$ increases linearly with $c_b$ for
$0.1M<c_b<1M$, as a consequence of a nearly constant $d$. In light
of these results, it is quite possible that the inhomogeneous
screening of the image force provides an explanation on the linear
increase of the surface tension with the salt concentration; we are
currently exploring this possibility.

\begin{figure}[hb]
\centering
\includegraphics[width=0.42\textwidth]{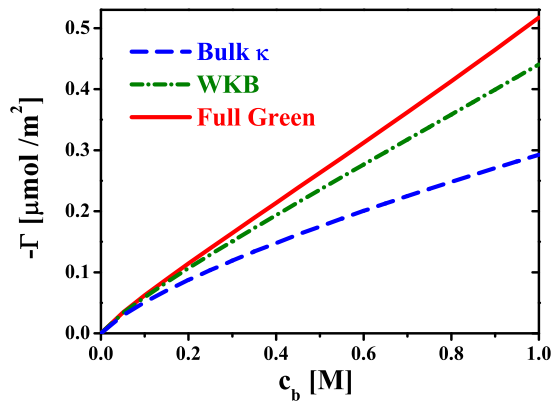}
\caption{Inhomogeneous screening effect on the negative
adsorption of ions for a 1:1 salt solution at the water/air interface.
 \label {2}}
\end{figure}

Inhomogeneous screening has an even more pronounced effect on asymmetric
salt solutions containing multivalent ions, because multivalent ions are more strongly depleted
than monovalent ions and are more effective in screening.  For a 2:1 salt solution,
the divalent cations calculated by fully solving the Green function are pushed further
away from the interface than predicted by the approximate methods as shown in Fig 5(a), leading to a larger degree of charge separation.  As a result,
the induced electrostatic potential is much larger than that obtained using the approximate methods; see Fig. 5(b).  Such a large self-induced
surface potential can significantly affect the interpretation of the
zeta potential of colloidal surfaces\cite{Israelachvili} and is a
major contribution to the Jones-Ray effect in the surface tension of
salt solution\cite{Onuki2,JonesRay,Bier,wang3}.

\begin{figure}[t]
\centering
\includegraphics[width=0.5\textwidth]{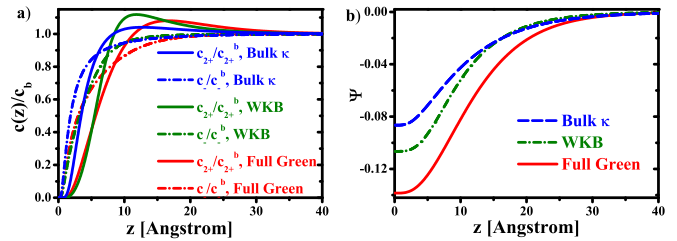}
\caption{Inhomogeneous screening effect on 0.05M 2:1 salt solution near the water/air
interface. (a) Ion concentration scaled by $c_{\pm}^b$ and (b)
dimensionless electrostatic potential.
 \label {2}}
\end{figure}

In conclusion, we have presented a self-consistent treatment of
the inhomogeneous screening in salt solutions near a dielectric interface.
The effect of inhomogeneous screening is twofold:  the lower ionic
strength near the interface results in less
screening on the image force and hence stronger ion depletion, and the depletion zone has a long-range and accumulative effect on screening, which extends the range
of the depletion layer.   Consequently, the ion distribution is
significantly affected when the bulk screening length is comparable to or smaller than the Bjerrum
length.  In this regime, the double layer structure and
the interfacial properties cannot be described by either the bulk screening approximation or the
WKB approximation.  The characteristic length of the
depletion layer scales with the Bjerrum length, resulting in a linear increase
of the negative
adsorption of ions with concentration, in agreement with experiments.
The inhomogeneous screening effect becomes is pronounced in the
less polar solvent and for the ions of higher valency.

Nonuniform ion distribution near a dielectric interface exists in
many colloidal and biophysical systems. An accurate treatment of
inhomogeneous screening is important to fully understand the role of
electrostatic interactions, which in turn is
necessary for evaluating the various nonelectrostatic contributions,
such as the cavity energy,
hydration, and dispersion
forces in these systems.  The relative importance of these
nonelectrostatic contributions, and even their existence,  have been
a subject of controversy\cite{Ninhamrev}.   By a more
accurate treatment of the inhomogeneous screening effect, we are in
a better position to understand some long-standing problems, such as
the specific ion effects and salt concentration effects in the
surface tension.

Acknowledgment is made to the donors of the American
Chemical Society Petroleum Research Fund for partial
support of this research.


\begin{thebibliography}{}
\bibitem{Debye}
P. Debye and E. H\"uckel, {\it Phys. Z.}, {\bf 24}, 185 (1923).
\bibitem{Israelachvili}
J. N. Israelachvili, {\it Intermolecular and Surface Forces}, 2nd
Ed. (Academic, London, 1992).
\bibitem{Andelmanreview}
D. Andelman, in {\it Soft Condensed Matter Physics in Molecular and
Cell Biology}, W. C. K. Poon and D. Andelman, eds. (Taylor and
Francis, Boca Raton, Florida, 2000).
\bibitem{McQuarrie}
D. A. McQuarrie, {\it Statistical Mechanics}, (University Science
Books, Sausalito, California, 2000).
\bibitem{Levinreview}
Y. Levin, {\it Rep. Prog. Phys.}, {\bf 65}, 1577 (2002).
\bibitem{Hansen}
J. P. Hansen and H. L\"owen, {\it Annual Rev. Phys. Chem.}, {\bf
51}, 209-242 (2000).
\bibitem{Honig}
B. Honig and A. Nicholls, {\it Science}, {\bf 268}, 1144 (1995). W.
\bibitem{Gelbart}
M. Gelbart, R. F. Bruinsma, P. A. Pincus, and V. A. Parsegian, {\it
Phys. Today}, {\bf 53}, 38 (2000).
\bibitem{onsager}
L. Onsager and N. N. T. Samaras, {\it J. Chem. Phys.}, {\bf 2}, 528
(1933).
\bibitem{wagner}
C. Wagner, Phys. Z. {\bf 25}, 474 (1924).
\bibitem{Channel1}
A. Parsegian, {\it Nature}, {\bf 221}, 844 (1969).
\bibitem{Channel2}
T. Bastug and S. Kuyucak, {\it Biophys. J.}, {\bf 84}, 2871 (2003).
\bibitem{Channel3}
S. Buyukdagli, M. Manghi and J. Palmeri, {\it Phys. Rev, Lett.},
{\bf 105}, 158103 (2010).
\bibitem{Channel4}
E. D. Gomez, A. Panday, E. H. Feng, V. Chen, G. M. Stone, A. M.
Minor, C. Kisielowski, K. H. Downing, O. Borodin, G. D. Smith, and
N. P. Balsara, {\it Nano Lett.}, {\bf 9}, 1212-1216 (2009).
\bibitem{Stability1}
Tavares, F. W., Bratko, D., Prausnitz, J. M. {\it Curr. Opin.
Colloid Interface Sci.} {\bf 9}, 81 (2004).
\bibitem{Stability2}
Gradzielski, M. {\it Curr. Opin. Colloid Interface Sci.} {\bf 9},
256 (2004).
\bibitem{Stability3}
V. S. J. Craig, B. W. Ninham and R. M. Pashley, {\it Nature} {\bf
364}, 317 (1993).
\bibitem{Stability4}
S. Kumar and R. Nussinov, {\it Chembiochem} {\bf 3}, 604 (2002).
\bibitem{Ozone1}
J. H. Hu, Q. Shi, P. Davidovits, D. R. Worsnop, M. S. Zahniser and
C. E. Kolb, {\it J. Phys. Chem.}, {\bf 99}, 8768 (1995).
\bibitem{Ozone2}
E. M. Knipping, M. J. Lakin, K. L. Foster, P. Jungwirth, D. J.
Tobias, R. B. Gerber, D. Dabdub, and B. J. Finlayson-Pitts, {\it
Science}, {\bf 288}, 301 (2000).
\bibitem{surfacetension1}
N. Matubayasi, K. Yamamoto, S. Yamaguchi, H. Matsuo and N. Ikeda,
{\it J. Colloid Interface Sci.}, {\bf 214}, 101 (1999).
\bibitem{surfacetension2}
N. Matubayasi, K. Tsunemoto, I. Sato, R. Akizuki, T. Morishita, A.
Matuzawa, and Y. Natsukari, {\it J. Colloid Interface Sci.}, {\bf
243}, 444 (2001).
\bibitem{Levin1}
Y. Levin and J. E. Flores-Mena, {\it Europhys. Lett.}, {\bf 56}, 187
(2001).
\bibitem{Levin2}
Y. Levin, A. P. dos Santos and A. Diehl, {\it Phys. Rev. Lett.},
 {\bf 103}, 257802 (2009).
\bibitem{Levin3}
A. Bakhshandeh, A. P. dos Santos and Y. Levin, {\it Phys. Rev.
Lett.}, {\bf 107}, 107801 (2011).
\bibitem{Monica}
J. W. Zwanikken and M. Olvera de la Cruz, {\it Proc. Natl. Acad.
Sci. USA}, {\bf 110}, 5301 (2013).
\bibitem{Bell2}
G. M. Bell and P. D. Rangecroft, {\it Trans. Faraday Soc.}, {\bf
67}, 649 (1971).
\bibitem{Ninham}
M. Bostr{\"o}m, W. Kunz and B. W. Ninham, {\it Langmuir}, {\bf 21},
2619 (2005).
\bibitem{Parsons}
T. T. Duignan, D. F. Parsons and B. W. Ninham, {\it J. Phys. Chem.
B}, {\bf 118}, 8700 (2014).
\bibitem{Onuki1}
A. Onuki, {\it Phys. Rev. E}, {\bf 73}, 021506 (2006).
\bibitem{Onuki2}
A. Onuki, {\it J. Chem. Phys.}, {\bf 128}, 224704 (2008).
\bibitem{Buyukdagli}
S. Buyukdagli, M. Manghi and J. Palmeri, {\it Phys. Rev. E}, {\bf
81}, 041601 (2010).
\bibitem{Dean}
D. S. Dean and R. R. Horgan, {\it Phys. Rev. E}, {\bf 69}, 061603
(2004).
\bibitem{Markovich}
T. Markovich, D. Andelman and R. Podgornik, {\it Europhys. Lett.},
{\bf 106}, 16002 (2014).
\bibitem{Wangzg}
Z. -G. Wang, {\it Phys. Rev. E}, {\bf 81}, 021501 (2010).
\bibitem{wang2}
R. Wang and Z. -G. Wang, {\it J. Chem. Phys.}, {\bf 142}, 104705
(2015).
\bibitem{boundary}
This boundary condition is obtained by combining the
continuity of ${\hat G}(k,z,z')$ at $z=0$ with the form of the analytical solution for
${\hat G}(k,z,z')$ in the $z<0$ region.
\bibitem{Orland}
R.R. Netz and H. Orland, {\it Eur. Phys. J. E},{\bf 11}, 301 (2003).
\bibitem{Buff}
F. P. Buff and F. H. Stillinger, {\it J. Chem. Phys.}, {\bf 39},
1911 (1963).
\bibitem{Levine}
G. M. Bell and S. Levine, {\it J. Chem. Phys.}, {\bf 49}, 4584
(1968).
\bibitem{wang1}
R. Wang and Z. -G. Wang, {\it J. Chem. Phys.}, {\bf 139}, 124702,
(2013).
\bibitem{numerical}
For each $k$, Eq. \ref{eq8} is solved with 2000 grid points for the
variable $z$ and 20000 grid points for the variable $z'$. We use
different discretization between $z$ and $z'$ to increase the
numerical accuracy for the self energy of the ions very close to the
interface. The Dirac delta function is approximated by the Kronecker
delta. Numerical integration in the $k$ space (Eq. \ref{eq11}) is
performed using the Simpson method with 200 grid points.
\bibitem{Jiaotong}
Z. L. Xu, M. M. Ma and P. Liu, {\it Phys. Rev. E}, {\bf 90}, 013307
(2014).
\bibitem{Fisher}
M. E. Fisher and Y. Levin, {\it Phys. Rev. Lett.}
 {\bf 71}, 3826 (1993).
\bibitem{JonesRay}
G. Jones and W. A. Ray, {\it J. Am. Chem. Soc.} {\bf 59}, 187
(1937).
\bibitem{Bier}
M. Bier, J. Zwanikken and R.van Roij, {\it Phys. Rev. Lett.} {\bf
101}, 046104 (2008).
\bibitem{wang3}
R. Wang and Z.-G. Wang, {\it J. Chem. Phys.} {\bf 135}, 014707
(2011).
\bibitem{cavity1}
K. Lum, D. Chandler, and J. D. Weeks, {\it J. Phys. Chem. B} {\bf
103}, 4570 (2005); D. Chandler, {\it Nature} {\bf 437}, 640 (2005).
\bibitem{cavity2}
S. Rajamani, T. M. Truskett, and S. Garde, {\it Proc. Natl. Acad.
Sci. U.S.A.} {\bf 102}, 9475 (2005).
\bibitem{dispersion}
M. Bostr{\"o}m, D. R. M. Williams and B. W. Ninham, {\it Langmuir}
{\bf 17}, 4475 (2001).
\bibitem{Ninhamrev}
P. L.  Nostro and B. W. Ninham, {\it Chem. Rev.} {\bf 112}, 2286
(2012).
\end{thebibliography}
\end{document}